\documentclass[aps,prl,twocolumn,showpacs,floatfix,superscriptaddress]{revtex4-1}

  \usepackage[utf8]{inputenc}
  \usepackage[T1]{fontenc}
  \usepackage{float} 
  \usepackage{color}  

\usepackage{graphicx}
\usepackage{amsmath}
\usepackage{amssymb}
\usepackage{bbm}
\usepackage{indentfirst}
\usepackage{dcolumn}
\usepackage{soul}

\begin{document}

\title{Skyrmions with attractive interactions in an ultrathin magnetic film}

\author{Levente R\'{o}zsa}
\email{rozsa.levente@wigner.mta.hu}
\affiliation{Institute for Solid State Physics and Optics, Wigner Research Centre for Physics, Hungarian Academy of Sciences,
P.O. Box 49, H-1525 Budapest, Hungary}
\author{Andr\'{a}s De\'{a}k}
\affiliation{Department of Theoretical Physics, Budapest University of Technology and Economics, Budafoki \'{u}t 8, H-1111 Budapest, Hungary}
\affiliation{MTA-BME Condensed Matter Research Group, Budapest University of Technology and Economics, Budafoki \'{u}t 8, H-1111 Budapest, Hungary}
\author{Eszter Simon}
\affiliation{Department of Theoretical Physics, Budapest University of Technology and Economics, Budafoki \'{u}t 8, H-1111 Budapest, Hungary}
\author{Rocio Yanes}
\affiliation{Department of Physics, University of Konstanz, D-78457 Konstanz, Germany}
\author{L\'{a}szl\'{o} Udvardi}
\author{L\'{a}szl\'{o} Szunyogh}
\affiliation{Department of Theoretical Physics, Budapest University of Technology and Economics, Budafoki \'{u}t 8, H-1111 Budapest, Hungary}
\affiliation{MTA-BME Condensed Matter Research Group, Budapest University of Technology and Economics, Budafoki \'{u}t 8, H-1111 Budapest, Hungary}
\author{Ulrich Nowak}
\affiliation{Department of Physics, University of Konstanz, D-78457 Konstanz, Germany}
\date{\today}

\begin{abstract}

We determined the parameters of a classical spin Hamiltonian describing an Fe monolayer on Pd$(111)$ surface with a Pt$_{1-x}$Ir$_{x}$ alloy overlayer from \textit{ab initio} calculations. While the ground state of the system is ferromagnetic for $x=0.00$, it becomes a spin spiral state as Ir is intermixed into the overlayer. Although the Dzyaloshinsky--Moriya interaction is present in the system, we will demonstrate that the frustrated isotropic exchange interactions play a prominent role in creating the spin spiral state, and these frustrated couplings lead to an attractive interaction between skyrmions at short distances. Using spin dynamics simulations, we show that under these conditions the individual skyrmions form clusters, and that these clusters remain stable at finite temperature.

\end{abstract}

\maketitle

The magnetic skyrmion corresponds to a configuration where the directions of the spin magnetic moments at different lattice sites span the whole sphere\cite{Nagaosa,Belavin}, in contrast to collinear ferromagnetic or antiferromagnetic systems and spin spiral states. Several years after the theoretical prediction\cite{Bogdanov,Bogdanov2}, a lattice of magnetic skyrmions has first been identified in the chiral magnet MnSi\cite{Muhlbauer}. Since this discovery, skyrmions have been detected experimentally in several other bulk systems; examples include FeGe\cite{Yu2,Wilhelm}, FeCoSi\cite{Munzer,Yu}, Cu$_{2}$OSeO$_{3}$\cite{Adams}, GaV$_{4}$S$_{8}$\cite{Kezsmarki}, and Co-Zn-Mn alloys\cite{Tokunaga}.

In agreement with the original theoretical description\cite{Bogdanov2,Bogdanov3}, the appearance of skyrmions in the above systems was attributed to the Dzyaloshinsky--Moriya interaction\cite{Dzyaloshinsky,Moriya} present in noncentrosymmetric magnets. This chiral interaction competes with the ferromagnetic exchange and easy-axis anisotropy, and may lead to a planar spin spiral ground state in the system\cite{Dzyaloshinsky2,Izyumov}, which can in turn transform into a skyrmion lattice at finite external magnetic field.

Since frustrated isotropic exchange interactions may also stabilize a spin spiral phase, skyrmions could also be present in such systems at finite external magnetic field, even if the Dzyaloshinsky--Moriya interaction is absent due to symmetry reasons. It was shown in Ref.~\cite{Okubo} for a model Hamiltonian with competing ferromagnetic and antiferromagnetic interactions on a triangular lattice that at least at finite temperature, this is indeed the case. It was demonstrated later\cite{Leonov,Lin,Hayami} that the presence of an easy-axis on-site anisotropy extends the stability range of the skyrmion lattice to zero temperature. If only isotropic exchange interactions are present, Bloch-type and N\'{e}el-type skyrmions with different helicities, as well as skyrmions and antiskyrmions with opposite topological charges\cite{Leonov}, are energetically degenerate. Furthermore, the magnetization profile of skyrmions with frustrated exchange interactions is different from that of skyrmions stabilized by the Dzyaloshinsky--Moriya interaction. This leads to an interaction potential between skyrmions with several local energy minima, while the interaction between Dzyaloshinsky--Moriya skyrmions is repulsive at all distances at low temperature\cite{Rossler}.

Magnetic skyrmions have also been explored in ultrathin film systems such as PdFe bilayer\cite{Romming} or Fe triple-layer\cite{Hsu} on Ir$(111)$ surface, and Pt|Co|Ir multilayers\cite{Moreau-Luchaire}. Since bulk inversion symmetry is broken at the surface, the Dzyaloshinsky--Moriya interaction is present in such systems; consequently, the theoretical descriptions\cite{Dupe,Simon,Hagemeister,Hagemeister2} so far have been based on the conventional model\cite{Bogdanov,Bogdanov2}. On the other hand, several recent publications\cite{Dupe,Rozsa,Dupe2} have identified the frustrated isotropic exchange interactions as the driving mechanism behind the creation of spin spiral ground states in specific ultrathin films.

Due to their size being in the nanometer regime and the fact that they can be manipulated by relatively weak spin-polarized currents\cite{Iwasaki,Jonietz}, skyrmions are promising candidates for future applications in data storage and logic devices\cite{Fert,Iwasaki2,Zhou}. At finite temperature, isolated skyrmions propagate diffusively on the field-polarized background\cite{Schutte}, and their uncontrolled motion leads to a loss of information in memory devices. It has been demonstrated in simulations\cite{Iwasaki,Muller} and experiments\cite{Woo} that it is possible to control this diffusive motion by lattice defects.

In this study, we have performed \textit{ab initio} calculations on a (Pt$_{1-x}$Ir$_{x}$)Fe bilayer system on Pd$(111)$ surface to determine the coupling coefficients in a classical Hamiltonian. We will demonstrate using Landau--Lifshitz--Gilbert\cite{Nowak} spin dynamics simulations that individual skyrmions may be stabilized in the collinear field-polarized state of the system under experimentally realizable external magnetic fields. The Dzyaloshinsky--Moriya interaction is responsible for determining the helicity of skyrmions, while the frustrated exchange interactions modify their shape, and lead to an oscillating skyrmion-skyrmion interaction potential. Our simulations evidence that the short-range attractive interaction pins the skyrmions next to each other, and the formed skyrmion clusters are resistant against diffusion processes at finite temperature.

The classical Hamiltonian describing the magnetic moments in the Fe layer reads
\begin{eqnarray}
H=\frac{1}{2}\sum_{i \ne j}\boldsymbol{S}_{i}\mathcal{J}_{ij} \boldsymbol{S}_{j}+\sum_{i}\boldsymbol{S}_{i}\mathcal{K} \boldsymbol{S}_{i}-\sum_{i}M\boldsymbol{S}_{i}\boldsymbol{B},\label{eqn1}
\end{eqnarray}
where the unit vectors $\boldsymbol{S}_{i}$ represent the spins, and $\boldsymbol{B}$ denotes the external magnetic field. The $\mathcal{J}_{ij}$ exchange coupling and the $\mathcal{K}$ on-site anisotropy tensors, as well as the $M$ magnetic moment in Eq.~(\ref{eqn1}) have been determined by combining the screened Korringa--Kohn--Rostoker method\cite{Szunyogh,Zeller} with the relativistic torque method\cite{Udvardi}. For the details of the calculations see the Supplemental Material\cite{supp}. The isotropic exchange interactions $J_{ij}=\frac{1}{3}\textrm{Tr}\mathcal{J}_{ij}$ represent scalar Heisenberg couplings between the spins; the antisymmetric parts of the coupling tensors $D^{\alpha}_{ij}=\frac{1}{2}\varepsilon^{\alpha\beta\gamma}\mathcal{J}^{\beta\gamma}_{ij}$ can be identified with the Dzyaloshinsky--Moriya vectors\cite{Simon}. In the sign convention of Eq.~(\ref{eqn1}), $J_{ij}<0$ describes ferromagnetic coupling between the spins, while $J_{ij}>0$ is antiferromagnetic.

\begin{figure}
\includegraphics[width=0.95\columnwidth]{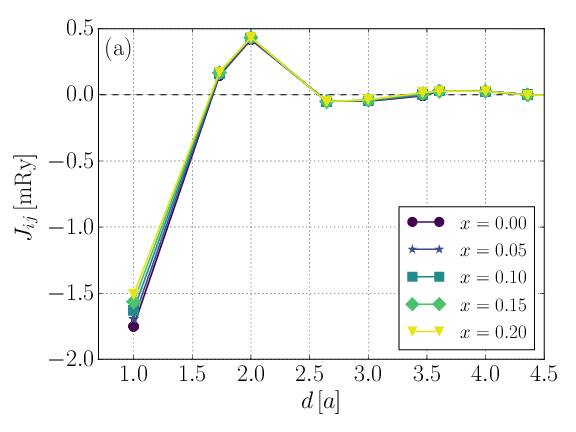}
\includegraphics[width=0.95\columnwidth]{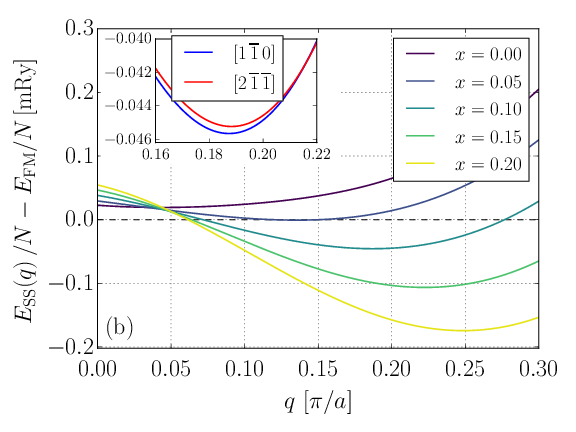}
\caption{(Color online) (a) Calculated isotropic exchange interactions $J_{ij}$ in (Pt$_{1-x}$Ir$_{x}$)Fe bilayer on Pd$(111)$ as a function of the distance $d$ between the Fe atoms, given in lattice constants of the triangular lattice $a$. (b) Energy per spin in the spin spiral state $E_{\textrm{SS}}\left(\boldsymbol{q}\right)/N$ as a function of the wave vector along the $[1\overline{1}0]$ direction, relative to the ferromagnetic state $E_{\textrm{FM}}/N$. Inset shows a slight anisotropy favoring the $[1\overline{1}0]$ axis over the $[2\overline{1}\overline{1}]$ direction for $x=0.10$.\label{fig1}}
\end{figure}

The $J_{ij}$ isotropic exchange interactions are depicted in Fig.~\ref{fig1}(a). Partially replacing Pt by Ir in the nonmagnetic overlayer decreases the magnitude of the nearest-neighbor ferromagnetic exchange interaction, while it does not influence the antiferromagnetic interactions with the second and third neighbors considerably. This means that decreasing the average number of valence electrons in the overlayer drives the system from the ferromagnetic towards the spin spiral state, in agreement with the results of Ref.~\cite{Dupe2} for a similar layered system.
% The difference compared to Ref.~\cite{Dupe2} is that we have used the $5d$ elements Pt and Ir instead of their $4d$ counterparts Pd and Rh.

%\begin{figure*}
%\includegraphics[width=1.9\columnwidth]{fig2.eps}
%\caption{Stable localized spin configurations with different topological charges: (a) skyrmion with $Q=-3$, (b) skyrmion with $Q=-2$\cite{Leonov,Lin}, (c) skyrmion with $Q=-1$, (d) ``chimera'' skyrmion with $Q=0$, (e) antiskyrmion with $Q=1$, and (f) antiskyrmion with $Q=2$\cite{Dupe3}. The exchange interactions for $x=0.05$ were used during the calculations. The value of the external field is $B=2.35\,\textrm{T}$ in part (a) and $B=0.23\,\textrm{T}$ in parts (b)-(f); the ground state is field-polarized for $B>0.21\,\textrm{T}$\cite{supp}.\label{fig3}}
%\end{figure*}

In order to determine the ground state of the system, we have calculated the energies $E_{\textrm{SS}}\left(\boldsymbol{q}\right)$ of harmonic spin spirals with wave vector $\boldsymbol{q}$, and compared them to the energy $E_{\textrm{FM}}$ of the ferromagnetic state along the easy out-of-plane direction. The results are summarized in Fig.~\ref{fig1}(b). For the calculations we have chosen right-handed cycloidal spin spirals,
\begin{eqnarray}
\boldsymbol{S}_{i}=\frac{\boldsymbol{q}}{\left|\boldsymbol{q}\right|}\sin\left(\boldsymbol{q}\boldsymbol{R}_{i}\right)+\boldsymbol{n}\cos\left(\boldsymbol{q}\boldsymbol{R}_{i}\right),\label{eqn1a}
\end{eqnarray}
where $\boldsymbol{n}$ is the outwards-pointing normal vector of the bilayer. Since the frustrated isotropic exchange interactions do not influence the rotational plane of the spiral, the energetically preferred right-handed cycloidal sense was determined by the Dzyaloshinsky--Moriya interactions, in agreement with the $C_{3\textrm{v}}$ symmetry of the surface\cite{Bogdanov2}. Due to the anisotropy of the system, $E_{\textrm{SS}}\left(\boldsymbol{q}\right)-E_{\textrm{FM}}$ does not converge to zero for harmonic spin spirals as $\boldsymbol{q}\rightarrow\boldsymbol{0}$, but it approximates the energy of the anharmonic spin spirals at finite wave vectors well\cite{Rozsa}.

Although the ground state of the system is out-of-plane ferromagnetic for a pure Pt overlayer $x=0.00$, the minimum of the spin spiral dispersion relation is below $E_{\textrm{FM}}$ at the higher Ir concentrations displayed in Fig.~\ref{fig1}(b). While the in-plane components of the Dzyaloshinsky--Moriya vectors prefer the creation of spin spiral states, they are weakening with increasing Ir concentration; for numerical values see the Supplemental Material\cite{supp}. This means that in the considered system the increasing frustration of the isotropic exchange interactions, shown in Fig.~\ref{fig1}(a), is responsible for the creation of the spin spiral. For small wave vectors, the anisotropy of the lattice only has a weak effect on the directional dependence of $E_{\textrm{SS}}\left(\boldsymbol{q}\right)$\cite{Hayami}. However, we note that spirals with wave vectors along $[1\overline{1}0]$ (the nearest neighbors in real space) are slightly preferred over ones with wave vectors along $[2\overline{1}\overline{1}]$ (the next-nearest neighbors), as shown in the inset of Fig.~\ref{fig1}(b).

\begin{figure}
\includegraphics[width=0.95\columnwidth]{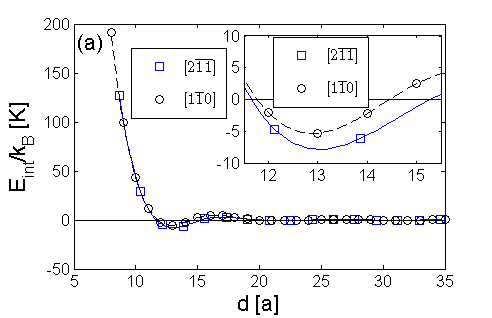}
\includegraphics[width=0.95\columnwidth]{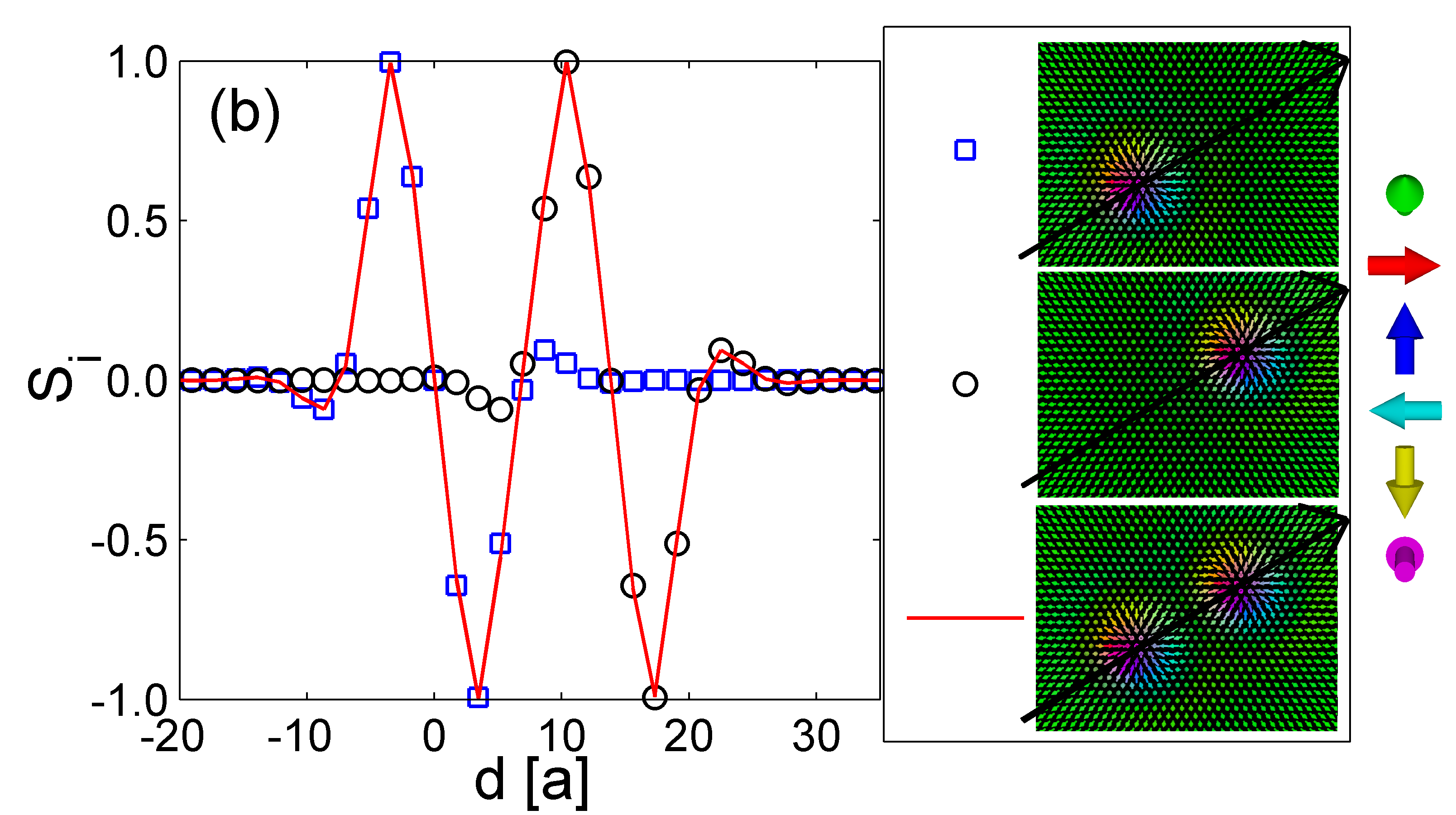}
\caption{(Color online) Attractive interaction between skyrmions for $x=0.10$ and $B=4.22\,\textrm{T}$. The field-polarized state represents the ground state for $B>4.18\,\textrm{T}$\cite{supp}. (a) Interaction energy between two skyrmions along the $[2\overline{1}\overline{1}]$ and $[1\overline{1}0]$ directions. Squares and circles denote data points, the interpolated lines are guides to the eye. Inset shows a close-up of the first minimum. (b) Real-space spin configuration of two isolated skyrmions (squares, circles) and two interacting skyrmions (line), with the centers of the skyrmions at the same lattice point in the two cases. $S_{i}$ is the projection of the vector $\boldsymbol{S}_{i}$ on the $[2\overline{1}\overline{1}]$ line connecting the centers, shown by arrows in the legend. The colorbar describes the orientation of the spins in the legend and in Fig.~\ref{fig3}.\label{fig2}}
\end{figure}

By applying an external magnetic field perpendicularly to the surface, the system will eventually transform into a collinear field-polarized state, possibly going through a skyrmion lattice phase for intermediate field values. We have observed localized noncollinear magnetic field configurations in the collinear phase by performing spin dynamics simulations. Due to the Dzyaloshinsky--Moriya interaction, skyrmions with topological charge $Q=-1$ are energetically the most favorable\cite{Nagaosa}, if the magnetization of the collinear state is pointing outwards from the surface. We calculated the interaction energy between two such skyrmions from numerical simulations, as illustrated in Fig.~\ref{fig2}. During the simulations, we fixed the spin at the center of the skyrmions to be antiparallel to the magnetization of the collinear state, and found the energy minimum with this constraint by the numerical solution of the Landau--Lifshitz--Gilbert equation. For $x\ge 0.05$, we found that the interaction energy oscillates while decaying. However, only the first local minimum is well visible in Fig.~\ref{fig2}(a) due to the exponential decay\cite{Leonov,Lin}. The presence of the local minima is clearly a consequence of the frustrated isotropic exchange interactions, since the Dzyaloshinsky--Moriya interaction prefers skyrmions that repulse each other at all distances; we found this to be the case for pure Pt overlayer ($x=0.00$), where the frustration of the interactions is the weakest. Note that the minimum is deeper for skyrmions separated along the $[2\overline{1}\overline{1}]$ direction compared to the $[1\overline{1}0]$ direction; this means that the preferred direction of nearest-neighbor bonds between skyrmions is perpendicular to the wave vector minimizing the spin spiral energies in Fig.~\ref{fig1}(b).
% Overall, this means that the hexagonal skyrmion lattice will be rotated by $\frac{\pi}{2}$ compared to the likewise hexagonal atomic lattice.

Figure~\ref{fig2}(b) demonstrates that the oscillation in the interaction energy is accompanied by an oscillation of the spin components in real space\cite{Leonov,Lin,Leonov3}. In the local minimum of the interaction potential, the skyrmions form a bond with the same sign of the in-plane spin component in the overlapping regime. On the other hand, the shape of skyrmions created by the Dzyaloshinsky--Moriya interaction can be well approximated by two domain walls located next to each other\cite{Romming2}, where the only sign change in the in-plane spin component is at the center of the skyrmion. The frustrated exchange interactions create further local extrema of the in-plane spin components where the rotational sense of the spins switches from right-handed to left-handed (helicity reversal\cite{Leonov}), which is energetically unfavorable from the standpoint of the Dzyaloshinsky--Moriya interactions. This means that the antiferromagnetic isotropic exchange interactions with the second and third neighbors are competing with not only the ferromagnetic nearest-neighbor interaction, but also with the Dzyaloshinsky--Moriya interactions, in order to form bonds between the skyrmions.

\begin{figure}
\includegraphics[width=0.90\columnwidth]{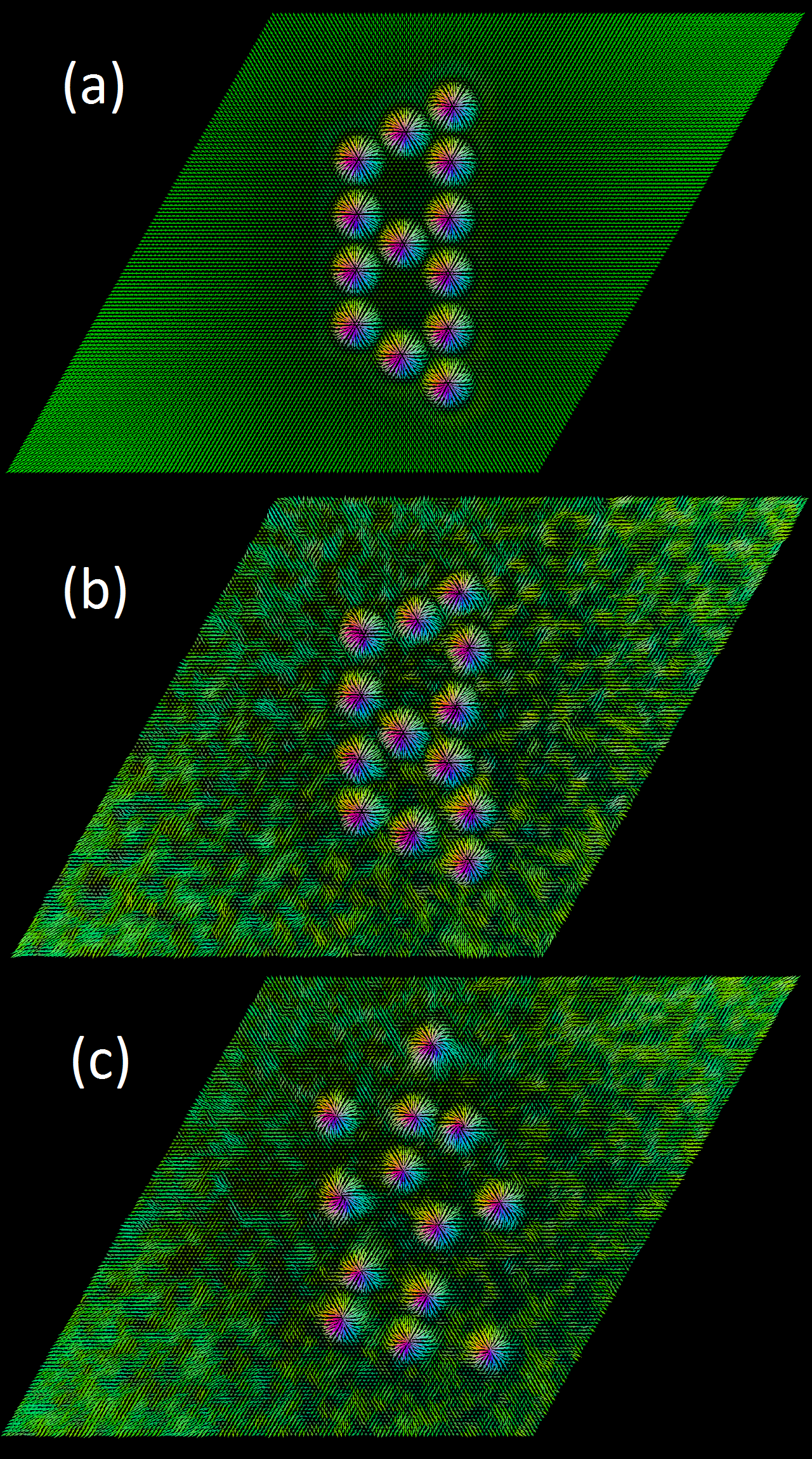}
\caption{(Color online) (a) Ordered initial configuration of skyrmions at $T=0\,\textrm{K}$, for $x=0.10$ and $B=4.22\,\textrm{T}$. (b)-(c) Final configuration after a thermalization of $t=605\,\textrm{ps}$ at $T=4.7\,\textrm{K}$, for different coupling parameters: (b) attractive skyrmions at $x=0.10$ and $B=4.22\,\textrm{T}$, (c) repulsive skyrmions at $x=0.00$ and $B=0.00\,\textrm{T}$. The lattice size is $N=128\times128$ atoms.\label{fig3}}
\end{figure}

Because the oscillating interaction potential determines an energetically favorable bond length between the skyrmions, it is possible to arrange them into arbitrarily shaped clusters at zero temperature. One example is displayed in Fig.~\ref{fig3}(a). As shown in Fig.~\ref{fig3}(b), the initial configuration is mostly conserved during simulations performed at $T=4.7\,\textrm{K}$, indicating thermal stability. For comparison, Fig.~\ref{fig3}(c) demonstrates how the information encoded in the original state is lost due to the diffusive motion and repulsive interaction between skyrmions with the system parameters $x=0.00$, $B=0.00\,\textrm{T}$.
%We have also examined how stable these arrangements are with respect to thermal fluctuations. At $T=4.7\,\textrm{K}$, skyrmions with attractive interactions in Fig.~\ref{fig3}(b) mostly conserve the initial configuration, while repulsive skyrmions in Fig.~\ref{fig3}(c) propagate diffusively in random directions, indicating loss of the information encoded in the original state.

\begin{figure}
\includegraphics[width=0.95\columnwidth]{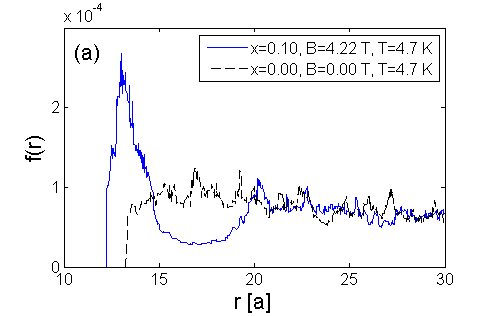}
\includegraphics[width=0.95\columnwidth]{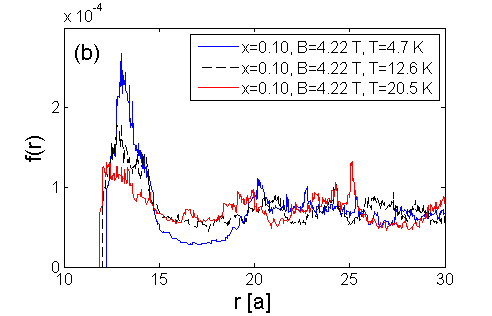}
\caption{(Color online) Pair correlation function $f\left(r\right)$ of skyrmions after a thermalization of $t=484\,\textrm{ps}$. (a) Difference between repulsive ($x=0.00$, $B=0.00\,\textrm{T}$) and attractive ($x=0.10$, $B=4.22\,\textrm{T}$) skyrmions at $T=4.7\,\textrm{K}$. (b) Temperature dependence of $f\left(r\right)$ for $x=0.10$ and $B=4.22\,\textrm{T}$. The system contained $31$ skyrmions on an $N=128\times128$ atomic lattice.\label{fig4}}
\end{figure}

The attractive interaction between skyrmions at finite temperature can also be characterized by calculating the pair correlation function $f\left(r\right)$, normalized as
\begin{eqnarray}
\int_{0}^{\infty}f\left(r\right)2\pi r\textrm{d}r=1.\label{eqn2}
\end{eqnarray}

%The method of calculating $f\left(r\right)$ is given in the Supplemental Material\cite{supp}. 
Figure~\ref{fig4} displays the pair correlation function after thermalization. We considered an initial configuration of $31$ skyrmions in random arrangement on an $N=128\times128$ lattice, the same size as in Fig.~\ref{fig3}; approximately $80$ skyrmions would fit into the same lattice size in a close-packed configuration with the applied simulation parameters. Figure~\ref{fig4}(a) displays that the distribution is basically uniform in space outside the strongly repulsive core for repulsive skyrmions ($x=0.00$, $B=0.00\,\textrm{T}$), indicating that the diffusive motion is dominating in this case. On the other hand, one can clearly identify a preferred nearest-neighbor distance for attractive skyrmions ($x=0.10$, $B=4.22\,\textrm{T}$) around $r\approx 13\,a$, coinciding with the potential energy minimum in Fig.~\ref{fig2}(a). This favors the formation of clusters, similarly to the artificially created one in Fig.~\ref{fig3}(b). The normalization of $f\left(r\right)$ is ensured by a decreased number of skyrmions in the ring between $15$-$20\,a$. It is shown in Fig.~\ref{fig4}(b) that when the temperature becomes slightly higher ($T\approx20\,\textrm{K}$) than the energy barrier protecting the local minimum in Fig.~\ref{fig2}(a) ($\left|E_{\textrm{int}}/k_{\textrm{B}}\right|\approx8\,\textrm{K}$), the peak in the distribution function disappears, and the clusters are destroyed by thermal fluctuations.

To summarize, we have demonstrated that isolated magnetic skyrmions may be stabilized in (Pt$_{1-x}$Ir$_{x}$)Fe bilayer on Pd$(111)$. The frustrated isotropic exchange interactions create an oscillating skyrmion--skyrmion interaction potential, at the expense of the Dzyaloshinsky--Moriya interactions which prefer repulsion at all distances. Due to the attractive interaction, the skyrmions may be arranged into clusters. The bonds between the skyrmions stabilize their relative positions at finite temperature, which may be important for future applications in memory devices.

\begin{acknowledgments}

The authors thank Bertrand Dup\'{e} and Elena Vedmedenko for enlightening discussions. Financial support for this work was provided by the Deutsche Forschungsgemeinschaft via SFB 767 ``Controlled Nanosystems: Interaction and Interfacing to the Macroscale'' and by the Hungarian Scientific Research Fund under projects No.~K115575 and No.~K115632.

\end{acknowledgments}

\end{document}

\title{Supplemental Material to \\Skyrmions with attractive interactions in an ultrathin magnetic film}

\author{Levente R\'{o}zsa}
\email{rozsa.levente@wigner.mta.hu}
\affiliation{Institute for Solid State Physics and Optics, Wigner Research Centre for Physics, Hungarian Academy of Sciences,
P.O. Box 49, H-1525 Budapest, Hungary}
\author{Andr\'{a}s De\'{a}k}
\affiliation{Department of Theoretical Physics, Budapest University of Technology and Economics, Budafoki \'{u}t 8, H-1111 Budapest, Hungary}
\affiliation{MTA-BME Condensed Matter Research Group, Budapest University of Technology and Economics, Budafoki \'{u}t 8, H-1111 Budapest, Hungary}
\author{Eszter Simon}
\affiliation{Department of Theoretical Physics, Budapest University of Technology and Economics, Budafoki \'{u}t 8, H-1111 Budapest, Hungary}
\author{Rocio Yanes}
\affiliation{Department of Physics, University of Konstanz, D-78457 Konstanz, Germany}
\author{L\'{a}szl\'{o} Udvardi}
\author{L\'{a}szl\'{o} Szunyogh}
\affiliation{Department of Theoretical Physics, Budapest University of Technology and Economics, Budafoki \'{u}t 8, H-1111 Budapest, Hungary}
\affiliation{MTA-BME Condensed Matter Research Group, Budapest University of Technology and Economics, Budafoki \'{u}t 8, H-1111 Budapest, Hungary}
\author{Ulrich Nowak}
\affiliation{Department of Physics, University of Konstanz, D-78457 Konstanz, Germany}
\date{\today}

\begin{abstract}

In the Supplemental Material, we discuss the details of the \textit{ab initio} calculations and the numerical simulations.

\end{abstract}

\maketitle

\section{\textit{Ab initio} calculations\label{secs1}}

The electronic structure of the system was determined by the screened Korringa--Kohn--Rostoker method\cite{Szunyogh,Zeller}. The calculations were performed within the local spin density approximation (LSDA) using the parametrization of the exchange-correlation potential in Ref.~\cite{Vosko}, and the atomic sphere approximation. We have checked that using the generalized gradient approximation (GGA) with the potential in Ref.~\cite{Perdew} instead of the LSDA reproduces the transition from the out-of-plane ferromagnetic to the spin spiral state with increasing Ir concentration; here we only present the interaction coefficients obtained using the LSDA. In the case of bulk Pd, we used the experimentally determined fcc lattice constant $a_{\textrm{Pd}}=3.891\,\textrm{\AA}$, being $\sqrt{2}$ times larger than the lattice constant $a$ of the triangular lattice on the $(111)$ surface. The surface structure consisted of ten layers of Pd, a single layer of Fe, an overlayer, and three layers of vacuum (empty spheres) located between the semi-infinite bulk and semi-infinite vacuum in fcc growth.

Lattice relaxations were included in the calculations between the Fe layer and the two neighboring layers. Their values were determined by VASP calculations\cite{Kresse,Kresse2,Hafner} using pseudopotentials from the projector-augmented wave method\cite{Blochl,Kresse3}, and an $11\times 11\times 1$ Monkhorst--Pack $\boldsymbol{k}$-mesh. For Pt/Fe/Pd$(111)$, we obtained a $10.4\%$ inward relaxation of the Fe layer and an $8.3\%$ inward relaxation of the Pt layer. The Wigner--Seitz radii of the Fe layer and the overlayer were modified in the screened Korringa--Kohn--Rostoker calculations according to the relaxations.

For describing the Pt$_{1-x}$Ir$_{x}$ alloy overlayer, we applied the coherent potential approximation. We did not modify the geometry compared to the Pt/Fe/Pd$(111)$ system, since the ratio of Ir was kept below $20\%$ and the difference between the lattice constants of Pt and Ir is around $2\%$. The spin magnetic moment of Fe was $M=3.3\,\mu_{\textrm{B}}$ within $1\%$ accuracy for all considered compositions; we found induced moments of $0.3\,\mu_{\textrm{B}},0.4\,\mu_{\textrm{B}},$ and $0.4\,\mu_{\textrm{B}}$ for the Pt, Ir, and top-layer Pd atoms, respectively.

The $\mathcal{J}_{ij}$ coupling coefficients and $\mathcal{K}$ on-site anisotropy tensor in the Hamiltonian (Eq.~(1) in the main text) were determined using the relativistic generalization\cite{Udvardi} of the torque method\cite{Liechtenstein}. The algorithm is based on calculating the energy costs of infinitesimal rotations around different ferromagnetic states. We considered the out-of-plane ferromagnetic state, which is the ground state of Pt/Fe/Pd$(111)$, and three nonparallel ferromagnetic orientations along the in-plane nearest-neighbor directions. If the coefficients are determined from these four calculations, they will reflect the $C_{3\textrm{v}}$ symmetry of the system. The energy integrations were performed on a semicircle contour containing $16$ energy points, while in the irreducible wedge of the Brillouin zone we included $1387$ $\boldsymbol{k}$ points at the first $8$ and $3104$ $\boldsymbol{k}$ points at the second $8$ energy points. The coupling coefficients were determined for neighbors within a radius of $8\,a$, netting a total of $240$ intralayer $\mathcal{J}_{ij}$ tensors, including those related to each other by symmetry.

%We note that the coupling coefficients are determined directly in real space, and therefore do not depend on the chosen radius -- this does not hold if they represent the result of a fitting procedure in reciprocal space\cite{Dupe,Ferriani}. However, one still has to check whether the decay of the coefficients is sufficiently fast, indicating that further neighbors can be omitted in the calculations. Fortunately, Fig.~1(a) in the main text and Fig.~\ref{figS1} demonstrate that the couplings beyond a radius of $3$-$4a$ are more or less negligible.

%Formally, it is possible to determine the coupling coefficients between the stable Fe moments and the induced moments in the neighboring layers. We performed these calculations to confirm that the induced moments have a minimal effect on the energetics of the spin system, and omitted these couplings from the simulations, since we do not expect that induced moments with fluctuating spin lengths may be described by a Hamiltonian of the form Eq.~(1).

\begin{figure}
\includegraphics[width=\columnwidth]{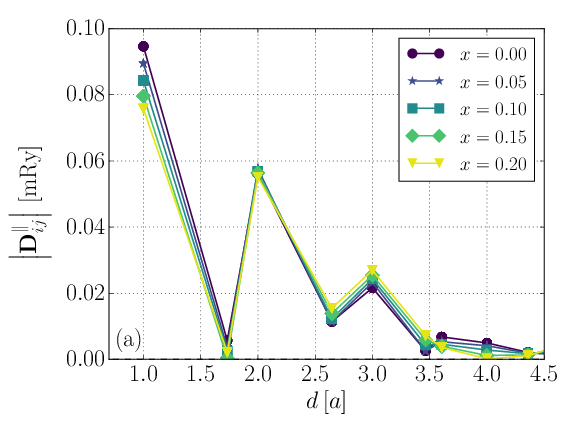}
\includegraphics[width=\columnwidth]{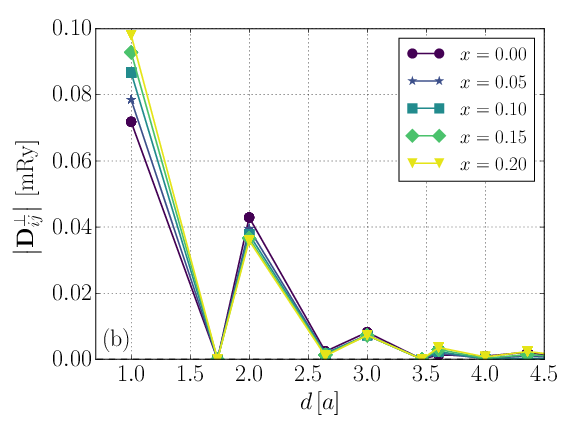}
\caption{(a) In-plane $\boldsymbol{D}_{ij}^{\Vert}$ and (b) out-of-plane $\boldsymbol{D}_{ij}^{\bot}$ components of the Dzyaloshinsky--Moriya vectors as a function of distance inside the Fe layer, for different Ir concentrations in the overlayer.\label{figS1}}
\end{figure}

To complement Fig.~1(a) in the main text, Fig.~\ref{figS1} presents the $\boldsymbol{D}_{ij}^{\Vert}$ in-plane and $\boldsymbol{D}_{ij}^{\bot}$ out-of-plane components of the Dzyaloshinsky--Moriya vectors between the Fe atoms. Note that for fcc$(111)$ surfaces, the presence of the out-of-plane components is only excluded by symmetry for specific pairs of atoms such as the next-nearest neighbors\cite{Crepieux}. Similarly to recent results for Co\cite{Yang} instead of Fe, we found that the in-plane component of the nearest-neighbor Dzyaloshinsky--Moriya vector changes sign when replacing Pt by Ir in the overlayer, indicated by the decrease of $\left|\boldsymbol{D}_{ij}^{\Vert}\right|$ with increasing $x$ in Fig.~\ref{figS1}(a). The accompanying increase in the out-of-plane component in  Fig.~\ref{figS1}(b) indicates that this sign change happens through a rotation of the Dzyaloshinsky--Moriya vector around the lattice vector connecting the nearest neighbors, while the magnitude of the vector is only minimally affected by $x$. However, only the in-plane component influences the energy of the cycloidal spin spiral ground state found at $x\ge 0.05$. The creation of the spin spiral ground state despite the decrease of the in-plane Dzyaloshinsky--Moriya vector components can be explained by the frustrated exchange interactions as mentioned in the main text.

The energy per spin in the right-handed cycloidal spin spiral state in Eq.~(2) reads
\begin{eqnarray}
\frac{1}{N}E_{\textrm{SS}}\left(\boldsymbol{q}\right)=&&\frac{1}{2}\sum_{\boldsymbol{R}_{j}-\boldsymbol{R}_{i}}\frac{1}{2}\left(\hat{\boldsymbol{q}}\mathcal{J}_{ij}\hat{\boldsymbol{q}}+\boldsymbol{n}\mathcal{J}_{ij}\boldsymbol{n}\right)\cos\left[\boldsymbol{q}\left(\boldsymbol{R}_{j}-\boldsymbol{R}_{i}\right)\right]\nonumber
\\
&&+\frac{1}{2}\sum_{\boldsymbol{R}_{j}-\boldsymbol{R}_{i}}\boldsymbol{D}_{ij}\left(\hat{\boldsymbol{q}}\times\boldsymbol{n}\right)\sin\left[\boldsymbol{q}\left(\boldsymbol{R}_{j}-\boldsymbol{R}_{i}\right)\right]\nonumber
\\
&&+\frac{1}{2}\left(\hat{\boldsymbol{q}}\mathcal{K}\hat{\boldsymbol{q}}+\boldsymbol{n}\mathcal{K}\boldsymbol{n}\right),\label{eqnS-1}
\end{eqnarray}
with $\hat{\boldsymbol{q}}=\boldsymbol{q}/\left|\boldsymbol{q}\right|$. In Fig.~1(b), this is compared to the energy of the out-of-plane ferromagnetic state,
\begin{eqnarray}
\frac{1}{N}E_{\textrm{FM}}\left(\boldsymbol{q}\right)=\frac{1}{2}\sum_{\boldsymbol{R}_{j}-\boldsymbol{R}_{i}}\boldsymbol{n}\mathcal{J}_{ij}\boldsymbol{n}+\boldsymbol{n}\mathcal{K}\boldsymbol{n}
\label{eqnS-2}
\end{eqnarray}

As mentioned in the main text, Eq.~(\ref{eqnS-1}) does not converge to Eq.~(\ref{eqnS-2}) as $\boldsymbol{q}\rightarrow\boldsymbol{0}$, because the on-site anisotropy energy in the harmonic spin spiral state does not depend on the magnitude of the wave vector $\boldsymbol{q}$, but differs from the value were all spins are parallel to the easy axis. The system gains anisotropy energy with respect to Eq.~(\ref{eqnS-1}) by forming an anharmonic equilibrium spin spiral state. We have confirmed with spin dynamics simulations that the ground state of the system is out-of-plane ferromagnetic for $x=0.00$ and a right-handed cycloidal spin spiral for $x\ge 0.05$ -- see Sec.~\ref{secs4} below. We note that the ground state of Fe/Pd$(111)$, that is the same system without the overlayer, is also ferromagnetic (cf. Ref.~\cite{Dupe}).

\section{Simulation methods\label{secs2}}

During the simulations, we have numerically solved the stochastic Landau--Lifshitz--Gilbert equation\cite{Nowak},
\begin{eqnarray}
\frac{\textrm{d}\boldsymbol{S}_{i}}{\textrm{d}t}&=&-\gamma' \boldsymbol{S}_{i} \times \left(\boldsymbol{B}_{i}^{\textrm{eff}}+\boldsymbol{B}_{i}^{\textrm{th}}\right) \nonumber
\\
&&- \gamma' \alpha\boldsymbol{S}_{i} \times\left[\boldsymbol{S}_{i} \times \left(\boldsymbol{B}_{i}^{\textrm{eff}}+\boldsymbol{B}_{i}^{\textrm{th}}\right) \right].\label{eqnS1}
\end{eqnarray}

The coupling coefficients enter the calculations through the effective field $\boldsymbol{B}_{i}^{\textrm{eff}}=-\frac{1}{M}\frac{\partial H}{\partial \boldsymbol{S}_{i}}$. The dimensionless Gilbert damping coefficient is denoted by $\alpha$, and $\gamma'=\frac{\gamma}{1+\alpha^{2}}$ corresponds to the modified gyromagnetic ratio with $\gamma=\frac{ge}{2m}$ ($g,e,m$ are the electronic spin $g$ factor, charge, and mass, respectively). The damping determines the speed of the relaxation to the nearest local energy minimum at zero temperature, and also the strength of the coupling to the heat bath through the thermal noise  $\boldsymbol{B}_{i}^{\textrm{th}}(t)=\sqrt{\frac{2\alpha k_{\textrm{B}}T}{M\gamma}}\circ\boldsymbol{\eta}_{i}(t)$. The $\circ$ symbol denotes the use of Stratonovich stochastic calculus in the interpretation of the stochastic differential equation~(\ref{eqnS1})\cite{Garcia-Palacios}. The numerical integrations were performed by the semi-implicit B method from Ref.~\cite{Mentink}.

For the calculation of the topological charge, we evaluated the spherical surface areas spanned by the spin vectors, which is the appropriate generalization of the continuum expression\cite{Belavin} for lattice models\cite{Berg}. For periodic boundary conditions used throughout the simulations, the topological charge will always be an integer within numerical accuracy\cite{Rozsa2}.

\section{Zero-temperature simulations\label{secs4}}

During the simulations we compared the energies of the cycloidal spin spiral, the hexagonal skyrmion lattice, and the field-polarized state along the out-of-plane direction as a function of external magnetic field. With the sets of interaction parameters used in this paper, we have not observed the other ordered phases discussed in Refs.~\cite{Leonov,Lin} for frustrated exchange interactions, either because the anisotropy was sufficiently large, or due to the presence of the Dzyaloshinsky--Moriya interactions. The equilibrium configuration was found after energy minimization at zero temperature according to Eq.~(\ref{eqnS1}), using the value $\alpha=1$ on an $N=128\times128$ lattice.

\begin{figure}
\includegraphics[width=\columnwidth]{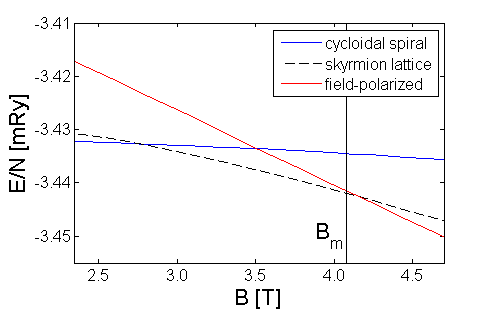}
\caption{Energy of different ordered phases as a function of external magnetic field, for $x=0.10$ on an $N=128\times128$ lattice. $B_{\textrm{m}}$ is the value of the external magnetic field where a single skyrmion on a field-polarized background becomes metastable.\label{figS2}}
\end{figure}

The results for $x=0.10$ are summarized in Fig.~\ref{figS2}. The wave vector of the spin spiral state was $q=0.094\frac{2\pi}{a}$ ($\lambda=2.9\,\textrm{nm}$) along the energetically favored $[1\overline{1}0]$ direction -- see Fig.~1(b). The skyrmions formed a triangular lattice, where the nearest neighbors were located along the $[2\overline{1}\overline{1}]$ directions of the atomic lattice -- see Fig.~2(a). The lattice consisted of $80$ skyrmions. The periodic boundary conditions did not allow a change in the wave vector of the spiral or the number of skyrmions. For larger values of the magnetic field, the skyrmions formed a cluster with a narrow field-polarized stripe at the periodic boundary due to the attractive interaction; however, even at the transition field from the skyrmion lattice to the field-polarized state $B=4.18\,\textrm{T}$, no additional skyrmion could fit into this field-polarized stripe.

Despite the limitations of using periodic boundary conditions, Fig.~\ref{figS2} gives a good approximation for the transition field values from the cycloidal spiral through the skyrmion lattice into the field-polarized state. Similarly to the case of repulsive skyrmions\cite{Dupe,Dupe2,Simon}, the phase diagram indicates that the magnetization of the system increases between the phases\cite{Bogdanov2}, and the subsequent states gain more energy from the Zeeman term in the Hamiltonian. The attractive interaction is indicated by the fact that a single skyrmion on a field-polarized background becomes energetically unfavorable at a lower field value ($B_{\textrm{m}}=4.08\,\textrm{T}$) than the transition point from the skyrmion lattice to the field-polarized state ($B=4.18\,\textrm{T}$).

In order to use isolated skyrmions for stable data storage or logic applications, the system must stay close to the magnetic field value $B_{\textrm{m}}$ where the presence of a skyrmion is energetically neutral; it has been demonstrated in Refs.~\cite{Schutte2,Hagemeister} that the lifetime of isolated skyrmions strongly depends on the value of the external field as well as the temperature. An absolute lower limit is given by the so-called elliptic instability field $B_{\textrm{e}}$\cite{Bogdanov3}, below which skyrmions strip out into spiral-like structures even at zero temperature.
% The field that collapses skyrmions on the lattice\cite{Leonov2} would represent an upper limit; however, in our simulations this value was found to be generally an order of magnitude larger than the other characteristic field values.

\begin{table}
\begin{ruledtabular}
    \begin{tabular}{rrrrr}
    $x$ & $B_{\textrm{e}}\left[\textrm{T}\right]$ & $B_{\textrm{m}}\left[\textrm{T}\right]$ & $B_{\textrm{s}}\left[\textrm{T}\right]$ & $\Delta E/N \left[\textrm{mRy}\right]$ \\
    0.00     & 0.00     & 0.00     & 0.00    & 0.0000 \\
    0.05  & 0.05 & 0.14 & 0.21 & 0.0029 \\
    0.10   & 3.05 & 4.08 & 3.46 & 0.0468 \\
    0.15  & 7.04 & 10.33 & 8.45 & 0.1074 \\
    0.20   & 12.67 & 17.37 & 13.61 & 0.1754 \\
    \end{tabular}%
\end{ruledtabular}
\caption{Characteristic magnetic field values for isolated skyrmions as a function of Ir concentration $x$. $B_{\textrm{e}}$ is the elliptic instability field; $B_{\textrm{m}}$ is the field value where an isolated skyrmion becomes metastable on the field-polarized background; $B_{\textrm{s}}$ is the field where the cycloidal spiral becomes metastable with respect to the field-polarized state; and $\Delta E/N$ is the energy difference between the out-of-plane ferromagnetic state and the ground state at $B=0\,\textrm{T}$.\label{tableS2}}
\end{table}

The field values $B_{\textrm{e}}$ and $B_{\textrm{m}}$ are summarized in Table~\ref{tableS2}. To establish a connection with Fig.~1(b) in the main text, we have also listed the energy difference $\Delta E/N$ between the out-of-plane ferromagnetic state and the ground state, which is an anharmonic spin spiral state for $x\ge0.05$. The Zeeman term makes the field-polarized state energetically preferable to the spin spiral state at the field value $B_{\textrm{s}}$, which increases with the energy difference.

As mentioned in the main text, for pure Pt overlayer $x=0.00$ the ground state is ferromagnetic even in the absence of external magnetic field. For $x=0.05$, the spin spiral ground state transforms directly into the field-polarized state at $B_{\textrm{s}}=0.21\,\textrm{T}$, which is higher than the value where an isolated skyrmion or a skyrmion lattice becomes metastable. The intermediate skyrmion lattice ground state may also be absent in conventional Dzyaloshinsky--Moriya systems where the skyrmions repulse each other\cite{Bogdanov3}; the ultrathin film system Pd(hcp)/Fe/Ir$(111)$ provides an example for this behavior\cite{Dupe}.

At $x\ge0.10$, $B_{\textrm{s}}$ is between $B_{\textrm{e}}$ and $B_{\textrm{m}}$, leading to the phase diagram illustrated in Fig.~\ref{figS2}. We note that the lower field boundary of the skyrmion lattice is below $B_{\textrm{e}}$ ($2.82\,\textrm{T}$ and $3.05\,\textrm{T}$); from the literature it is known that this may also happen in the case of conventional repulsive skyrmions\cite{Bogdanov3}. Nevertheless, skyrmion lattices below $B_{\textrm{e}}$ are impractical for applications, since deleting a single skyrmion from the lattice invokes the strip-out instability in the neighboring ones.

\begin{figure*}
\includegraphics[width=2\columnwidth]{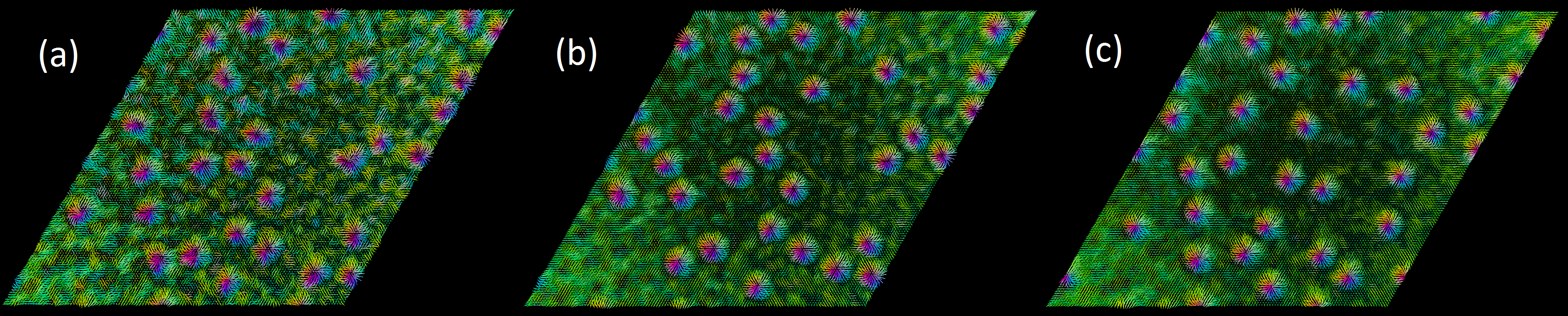}
\caption{(a) Initial configuration for the calculation of the pair correlation function, obtained after cooling down from the paramagnetic state until $T=15.8\,\textrm{K}$, for $x=0.10$ and $B=4.22\,\textrm{T}$. (b)-(c) Examples of final configurations after a thermalization of $t=484\,\textrm{ps}$ at $T=4.7\,\textrm{K}$, for different coupling parameters: (b) attractive skyrmions at $x=0.10$ and $B=4.22\,\textrm{T}$, (c) repulsive skyrmions at $x=0.00$ and $B=0.00\,\textrm{T}$. The lattice size is $N=128\times128$ atoms.\label{figS6}}
\end{figure*}

For calculating the strength of the interaction between skyrmions in Fig.~2(a), we initialized two skyrmion cores containing $5\times5$ spins on the field-polarized background, fixed the spin directions in the middle of the cores to be antiparallel to the direction of the external field, and found the equilibrium state by relaxation as before. The lattice size was $N=128\times128$ with periodic boundary conditions; since the interaction strength is negligible after a distance of $20$-$25$ atoms (see Fig.~2), the boundaries should have a minimal effect. The interaction energy is defined as
\begin{eqnarray}
E_{\textrm{int}}(d)=E_{\textrm{2sk}}(d)-2E_{\textrm{sk}},\label{eqnS5}
\end{eqnarray}
where $E_{\textrm{2sk}}$ is the energy of two interacting skyrmions and $E_{\textrm{sk}}$ is the energy of a single isolated skyrmion. All energies are measured with respect to the field-polarized state.

\section{Finite-temperature simulations\label{secs6}}

%\begin{figure}
%\includegraphics[width=\columnwidth]{figS3.eps}
%\caption{Pair correlation function of skyrmions after thermalization for $x=0.10$ and $B=4.22\,\textrm{T}$ as a function of temperature. The distribution was calculated from $10$ sample paths starting from the same initial configuration containing $31$ skyrmions on an $N=128\times128$ lattice.\label{figS5}}
%\end{figure}

%\begin{figure}
%\includegraphics[width=\columnwidth]{sk2distvarnew.eps}
%\caption{ Variance of the distance between two skyrmions as a function of simulation time, calculated from $50$ sample paths starting from the same initial configuration. The skyrmions repulse each other for $x=0.00$, $B=0.00\,\textrm{T}$, while the interaction energy is oscillating for $x=0.10$, $B=4.22\,\textrm{T}$ (Fig.~3(a)).\label{figS4a}}
%\end{figure}

The pair correlation function $f\left(r\right)$ corresponds to the radial probability density function of the distribution of the distances between the skyrmions. To calculate these distances at finite temperature, it is necessary to identify the location of larger objects in the atomic spin configuration $\left\{\boldsymbol{S}_{i}\right\}$. For this purpose, we interpolated the lattice spins on a rectangular grid, and found the skyrmions by template matching\cite{Bradski}, where the template was the isolated skyrmion at zero temperature. The similarity between the template and the image was quantified by calculating the correlation coefficient, using a scalar product for the three-dimensional spin vectors. At sufficiently low temperature, this method identified the same number of skyrmions that was determined from the topological charge during the simulations. The initial configuration only contained skyrmions with the same topological charge, and the temperature was significantly lower than where skyrmion creation and annihilation processes are observable under simulation timescales\cite{Rozsa2}.

For calculating the pair correlation function for an ensemble of skyrmions shown in Fig.~4 in the main text, we used an initial configuration containing $31$ skyrmions on an $N=128\times128$ lattice, obtained from cooling down the system from the paramagnetic state until $T=15.8\,\textrm{K}$ for $x=0.10$ and $B=4.22\,\textrm{T}$. This initial configuration is displayed in Fig.~\ref{figS6}(a). As it was discussed in Sec.~\ref{secs4}, approximately $80$ skyrmions fit into this lattice size in the close-packed skyrmion lattice state, thus the field-polarized state was less than half-filled with skyrmions in the initial configuration. We performed simulations for $x=0.10$, $B=4.22\,\textrm{T}$ and $x=0.00$, $B=0.00\,\textrm{T}$ with $10$ independent seeds each, and calculated all possible distances between skyrmions after a thermalization of $484\,\textrm{ps}$. Examples for these final configurations are shown in Figs.~\ref{figS6}(b)-(c). We have used a high value of the Gilbert damping $\alpha=1$, which increases the speed of diffusion processes\cite{Schutte}.
%besides providing the fastest relaxation, also maximizes the speed of diffusion processes.

As indicated by Figs.~\ref{figS6}(b)-(c), it is not always easy to determine just by looking at the real-space arrangement of skyrmions whether the interaction between them is attractive or repulsive, because the repulsion is weak, and skyrmions may randomly move next to each other during the diffusive motion. However, calculating the pair correlation function as in Fig.~4(a) reveals the difference between the two cases. Due to the finite size of the lattice, the pair correlation function always reached zero around $70$ lattice constants, since the periodic boundary conditions did not allow for larger distances; however, this value should not significantly influence the function in the $12$-$20\,a$ region, where the difference between repulsive and attractive skyrmions, as well as between attractive skyrmions at different temperatures, is the most prominent.